\documentclass[aps,prb,reprint,superscriptaddress,10pt]{revtex4-1}

\usepackage[]{graphicx}
\usepackage[utf8]{inputenc}
\usepackage[T1]{fontenc}%
\usepackage{dcolumn}%
\usepackage{bm}%
\usepackage{lmodern}%
\usepackage{amsmath, amssymb}
\usepackage{subfigure}%
\usepackage{hyperref}%
\usepackage{xcolor}%

\newcommand{\beq}{\begin{equation}\begin{aligned}}
\newcommand{\eeq}{\end{aligned}\end{equation}}

\newcommand{\CeCuSi}{CeCu$_{2}$Si$_{2}$}
\definecolor{linkcol}{rgb}{0,0,0.4}
\definecolor{citecol}{rgb}{0.5,0,0}

\hypersetup{
bookmarksopen=true,
pdfauthor="G.Giriat",
pdfsubject="Multiprobe CeCu2Si2", 
pdfstartview={FitH},    
pdfmenubar=true, 
pdfhighlight=/O, 
colorlinks=true, 
pdfpagemode=UseNone, 
pdfpagelayout=SinglePage, 
pdffitwindow=true, 
linkcolor=linkcol, 
citecolor=linkcol, 
urlcolor=blue
}

\begin{document}


\title{High pressure investigation of superconducting signatures in \CeCuSi{} : ac- magnetic susceptibility and heat capacity, resistivity and thermopower} 


%
\author{Gaétan Giriat}
\email{gaetan.giriat@unige.ch}
\author{Zhi Ren}
\affiliation{DQMP, Université de Genève, 24 quai Ernest Ansermet, CH-1211, Geneva, Switzerland}
\author{Pablo Pedrazzini}
\affiliation{Laboratorio de Bajas Temperaturas \& Instituto Balseiro, Centro Atómico Bariloche (CNEA), 8400 San Carlos de Bariloche, Argentina}
\author{Didier Jaccard}
\affiliation{DQMP, Université de Genève, 24 quai Ernest Ansermet, CH-1211, Geneva, Switzerland}


\date{\today}

\begin{abstract}
Taking advantage of a novel multiprobe setup we have measured, on a unique sample, the ac-magnetic susceptibility, the resistivity, the ac-specific heat and the thermopower of the superconductor heavy fermion \CeCuSi{} under pressure up to 5.1\,GPa. At the superconducting transition temperature $T_{\rm c}$, the Meissner signal corresponds to that expected for the sample volume and coincides with the specific heat jump and the resistive transition completion temperatures. Differing from previous observations, here the susceptibility measurements did not reveal any anomaly in the vicinity of the resistive transition onset.
\end{abstract}

\pacs{XXX}

\maketitle

\section{Introduction}
\label{}
Despite having been discovered in 1979\cite{Steglich1979}, the first unconventional superconductor \CeCuSi{} is still under investigation\cite{Kittaka2014} and the origin of its electronic properties remains controversial. At ambient pressure, a heavy fermion (HF) state develops on cooling, and superconductivity (SC), located in close proximity of a magnetic quantum critical point, is usually considered to be mediated by critical spin fluctuations, as it is the case for other HF superconductors\cite{Stockert2011,Monthoux2007,Lengyel2009,Lengyel2011}. At higher pressure ($p$) another regime appears, and around $p_V=4.5$\,GPa, valence (or charge) fluctuations associated with the critical end point of the valence transition line of Ce 4$f$ electrons, are believed to provide the glue for Cooper pairs\cite{Jaccard1999,Yuan2003,Holmes2004,Seyfarth2012,Miyake2007}. However, both hypotheses have recently been challenged. A new  thermodynamic study\cite{Kittaka2014} (at $p=0$) questions the spin mediated origin of the low $p$ SC and a new proposal suggests that orbital fluctuations are responsible for the pairing at high $p$.\\ \indent 
On pressure increase the superconducting transition temperature $T_{\rm c}\sim\,0.7$\,K  remains nearly unchanged up to 1-2\,GPa, but above that, $T_{\rm c}$ is sharply enhanced and reaches a maximum of $\sim$\,2.4\,K around $p_V$ before vanishing to zero. This non-monotonic trend of $T_{\rm c}(p)$ has been essentially probed by electrical resistivity measurements which invariably exhibit a dramatic broadening of the superconducting transition in an intermediate pressure range (1.5-3\,GPa) independently of the pressure transmitting medium hydrostaticity (He, daphne oil or steatite) and the sample quality as defined by the residual resistivity\cite{Bellarbi1984,Holmes2005}. Unexpectedly, the first ac-magnetic susceptibility measurements performed in an He-filled diamond anvil cell\cite{Thomas1996} point towards a $T_{\rm c}(p)$ dependence which would approximately follow the onset of the resistive transition $T_{\rm c}^{\rm onset}(p)$. It implies a discrepancy of up to 1\,K between the magnetic $T_{\rm c}$ and the completion of the resistive transition $T_{\rm c}^{\rm R=0}$ at pressures around 2.5\,GPa. In order to elucidate this issue, we decided to investigate four electronic properties (resistivity ($\rho$), thermopower ($S$), ac-heat capacity ($C_{\rm ac}$) and ac-magnetic susceptibility ($\chi_{\rm ac}$)) of a unique \CeCuSi{} sample with a high-$p$ multiprobe setup. From our measurements, the ac-magnetic susceptibility superconducting transition coincides with the jump in ac-heat capacity and with $T_{\rm c}^{\rm R=0}$ reflecting the material bulk property. Moreover the $\chi_{\rm ac}$ remains smooth near the resistivity $T_{\rm c}^{\rm onset}$ temperature.
\section{Methods}
\begin{figure}[t]
\centering
\includegraphics[width=.48\textwidth]{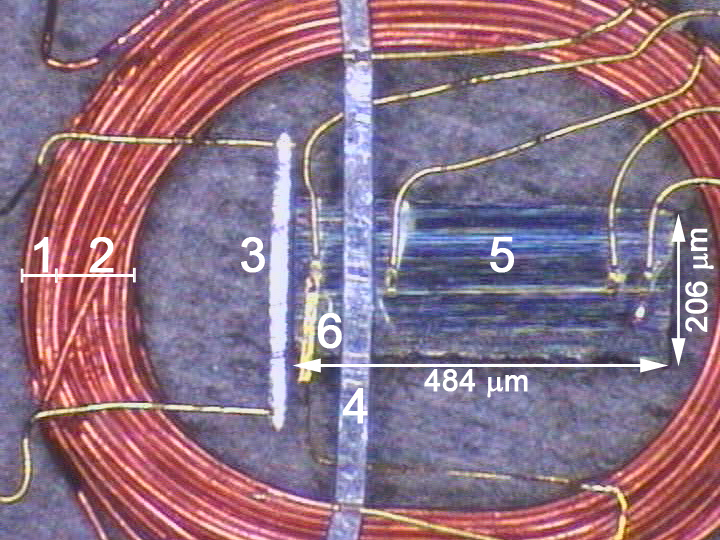}
\caption{Photo of the multiprobe setup sitting at the center of a non-magnetic tungsten carbide anvil flat. The 8 turns primary (1) and the 36 turns secondary (2), 14 $\mu$m insulated Cu wire coils are coaxial with a height of 95 $\mu$m. Au wires of 10 $\mu$m in diameter are spot welded on the chromel heater (3), the pressure calibrant Pb (4) and the sample \CeCuSi{} (5) whose c-axis is perpendicular to the picture plane. Near the heater, a 12 $\mu$m \underline{Au}Fe (0.07\,\% Fe) wire (6) adjoining a Au lead form the thermocouple used for ac-heat capacity or dc-thermopower measurements. Note: Two Au contacts located at each extremities of Pb are not shown on this picture.}
\label{fig:Setup}
\end{figure}

The modified Bridgman-anvil pressure cell employed for this experiment\cite{Ruetschi2007} accepts liquid pressure mediums thus providing a good hydrostaticity\cite{Klotz2009,Tateiwa2009} and a large working volume which are the required conditions to implement a multiprobe setup\cite{Seyfarth2012,Jaccard2010}. The highlight of this technique is that various physical properties of a unique sample are investigated under identical pressure conditions allowing accurate comparison of the results (insignificant differences might exist between the various regions of the sample being examined by each probe). Here the geometrical arrangement of the assembly together with the sample dimensions, depicted in Fig.\,\ref{fig:Setup}, result from a compromise between the constraints associated with each type of probe with a priority given to $\chi_{\rm ac}$ measurements. The setup developed for this experiment is placed at the centre of a pyrophyllite gasket with an initial internal diameter and thickness of 1.8 and 0.185\,mm respectively which is then filled with Daphne oil 7474 pressure medium\cite{Idemi} and sandwiched between two non-magnetic tungsten carbide anvils with 3.5\,mm flats.\\ \indent
The orientation of the magnetic coils, in the plane of the pressure cell, departs from our previous magnetic probe system\cite{Jaccard2010} and offers the possibility to examine much larger samples. The \CeCuSi{} specimen, $480\times210\times48\,\mu m^{3}$ (same batch as in Ref.\,\onlinecite{Vargoz1998}) is sufficiently large to generate a well resolved magnetic signal at $T_{\rm c}$, and remains compatible with the $\rho$, $C_{\rm ac}$ and $S$ measurement techniques described elsewhere\cite{Seyfarth2012,Jaccard2010}. An ac-excitation current of 1\,mA at 707\,Hz in the primary coil, produces an excitation field of 0.13 gauss and, although the ratio between the sample and the probing volume (filling factor) is 10.5\,\% in the present setup, the sample susceptibility drop to $\rho=-1$ at $T_{\rm c}$ induces an EMF of $\sim$\,60\,nV, well above the detection limit of standard apparatus. The joule heating of the primary coil is negligible and does not interfere with investigations down to 0.1\,K. With a signal-to-noise ratio of 103, we could also establish the superconducting transition temperature of the Pb manometer and compare it with that of the resistive transition for a better pressure calibration.\\ \indent
A comparison of the sample volume to its Meissner effect amplitude ratio with that calculated for the Pb shows that the field is expulsed by the whole sample with an accuracy of 5\,\%. This argument is well supported by calculations\cite{Alireza2003} based on the setup parameters which result in an induced EMF of $\sim$\,105\,nV. The 40\% difference between the calculated and measured value (60\,nV) falls into the incertitude on the coils dimensions and on the field strength. A difference of the same order exists between the expected and experimental value for the Pb.\\ \indent
\section{Results}
\begin{figure}[t]
\centering
\includegraphics[width=.48\textwidth]{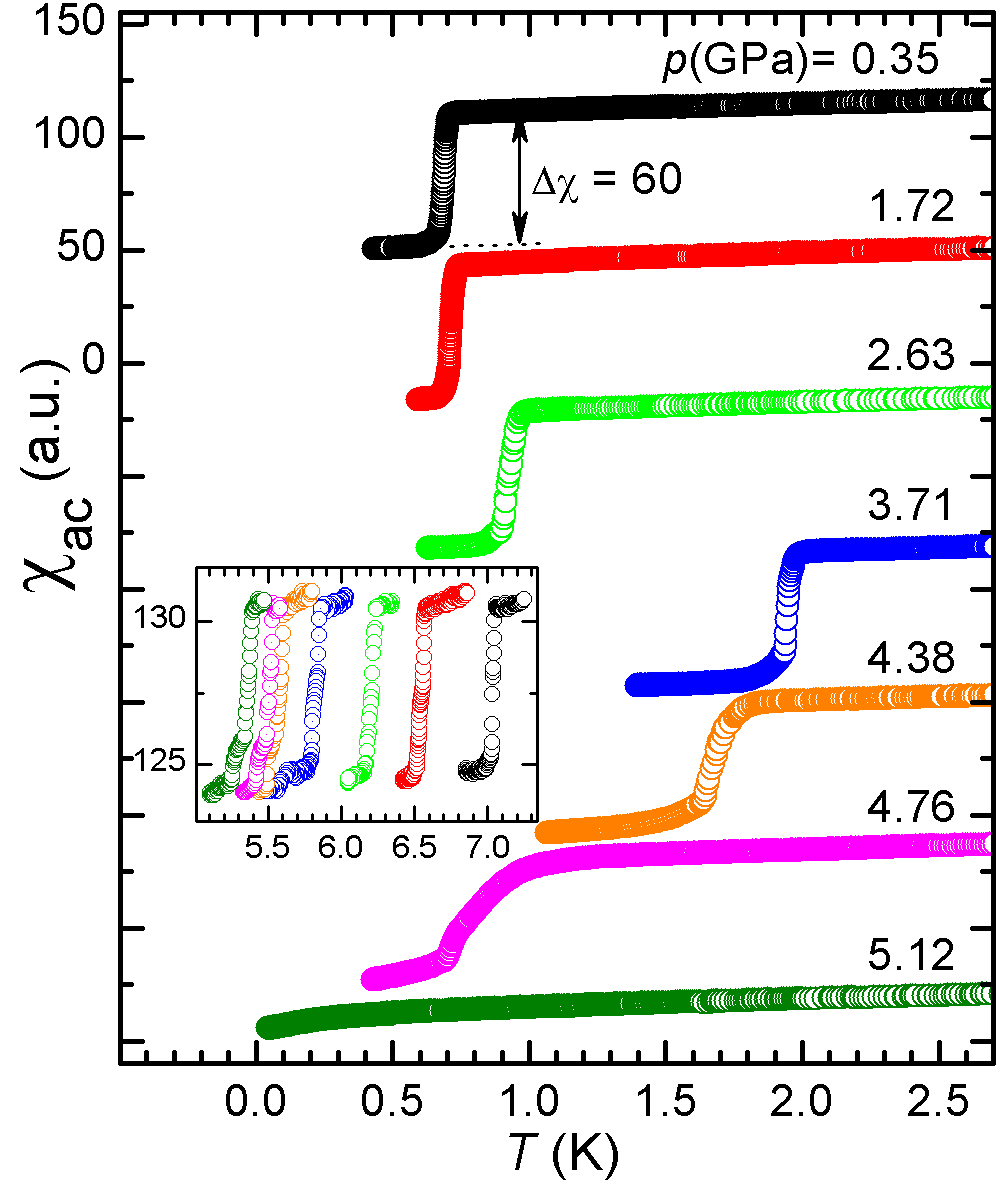}
\caption{\CeCuSi{} magnetic susceptibility $\chi_{\rm ac}$ in the proximity of $T_{\rm c}$ at various pressures. The magnetic susceptibility is reported in arbitrary units, however the values reported on the ordinate correspond to the voltage measured in nanovolts for the 0.35\,GPa data. Subsequent data sets are shifted down by a constant for clarity. The inset shows the lead $T_{\rm c}$ measured with identical settings.}
\label{fig:ACSusc_ALL}
\end{figure}

\begin{figure}[t]
\centering
\includegraphics[width=.48\textwidth]{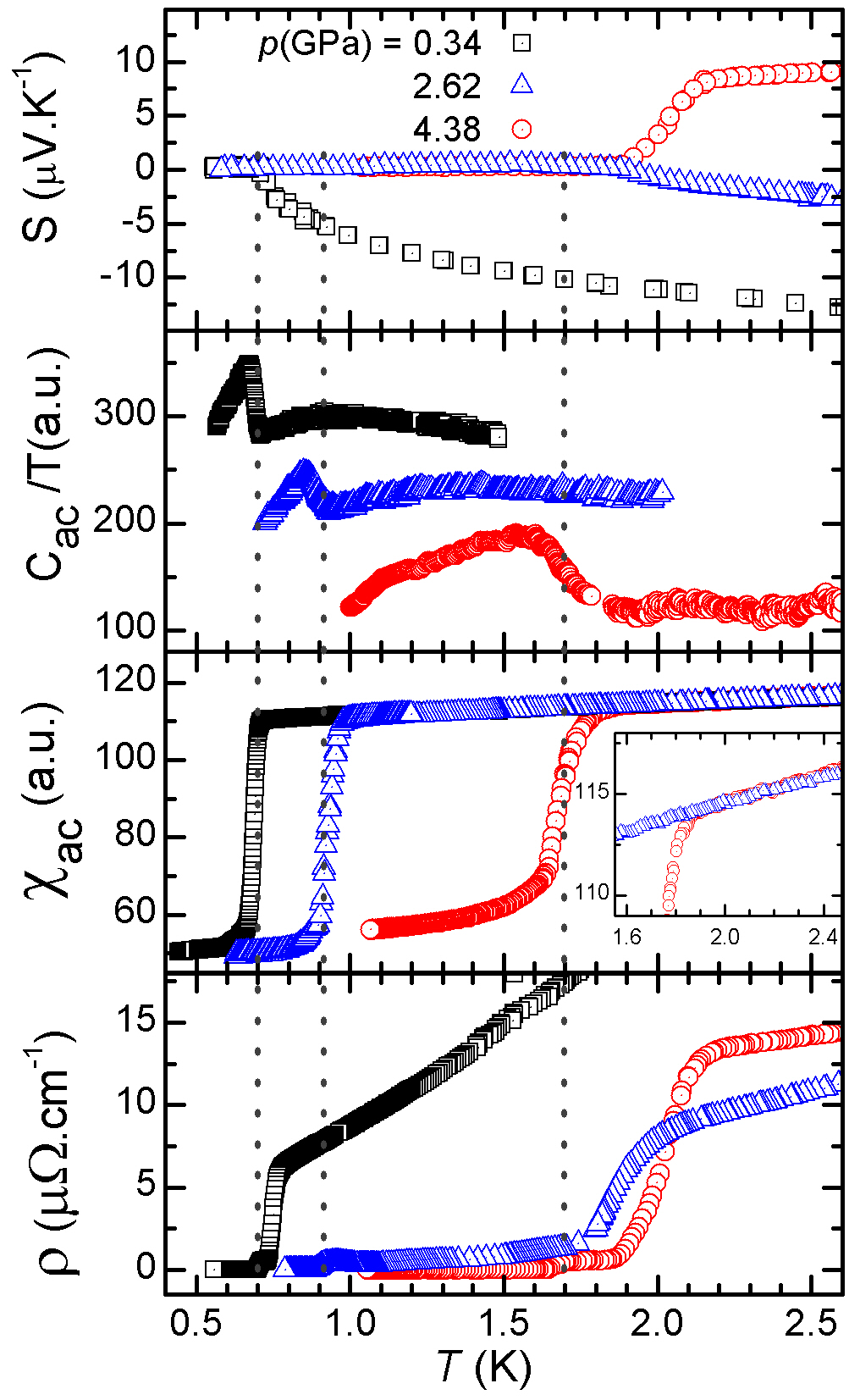}
\caption{Electronic properties under study at 0.34, 2.62 and 4.38\,GPa. Thermopower $S$, ac-heat capacity $C_{\rm ac}$, ac-magnetic susceptibility $\chi_{\rm ac}$ and resistivity $\rho$. The ac-heat capacity data at 0.34\,GPa was recorded at a different frequency and has been normalised. The reported ac-magnetic susceptibility is expressed in terms of induced voltage and the scale is in nanovolts. The $\chi_{\rm ac}$ inset illustrates the absence of anomaly in the temperature range 1.6-2.4\,K where the resistive transition onset for $p=2.62$ and 4.38\,GPa are located and provides also information on the noise level. The vertical dotted lines indicate bulk $T_{\rm c}$s at 0.34, 2.62 and 4.38\,GPa and are defined by the $C_{\rm ac}$ and $\chi_{\rm ac}$ transitions mid-points.}
\label{fig:TEP_CAC_Susc_Res}
\end{figure}

\begin{figure}[t]
\centering
\includegraphics[width=.48\textwidth]{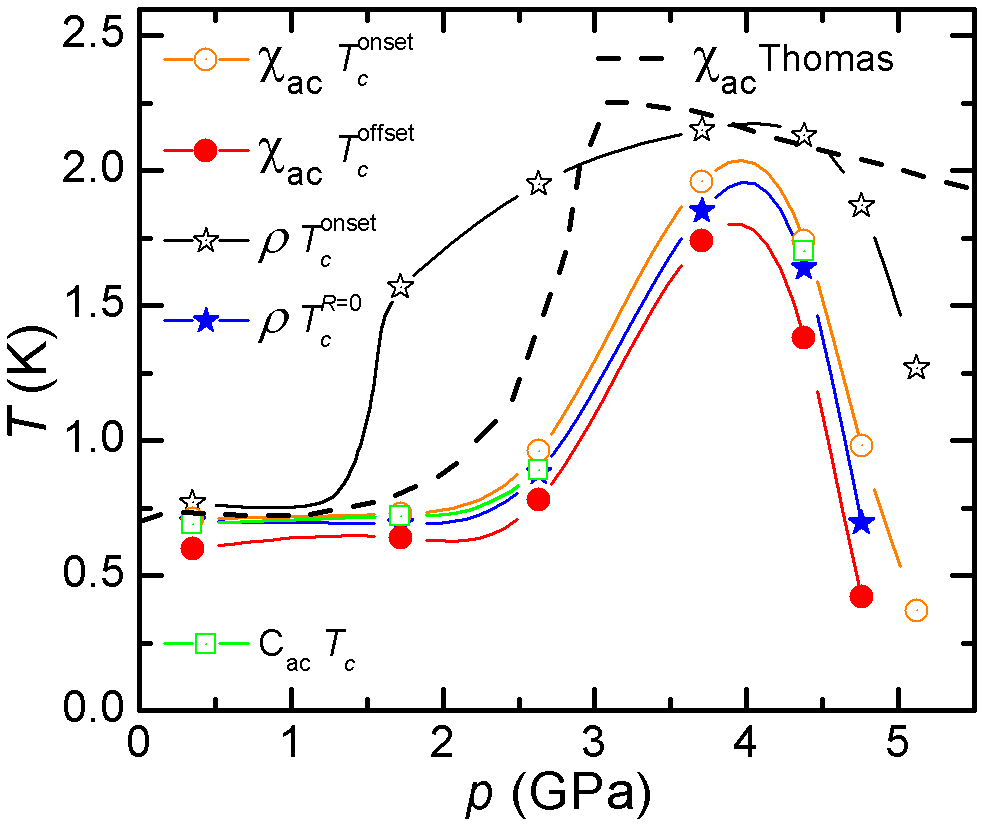}
\caption{$p\!-\!T$ phase diagram of \CeCuSi{} with $T_{\rm c}$ established through ac-heat capacity $C_{\rm ac}$, ac-magnetic susceptibility $\chi_{\rm ac}$ and resistivity $\rho$ measurements. Due to instability in the thermocouple gold wire contact, $C_{\rm ac}$ data could not be collected at 3.63\,GPa and 4.76\,GPa.}
\label{fig:Tc_vs_P}
\end{figure}

Susceptibility curves, of the \CeCuSi{} sample at different pressures are plotted in Fig.\,\ref{fig:ACSusc_ALL}. The data display a unique distinct anomaly superimposed over a linear temperature dependent background. The superconducting transition $T_{\rm c}$ comes to a maximum of 2\,K at 4\,GPa. Its width remains constant up to 2.65\,GPa but tends to broaden gradually in the subsequent runs. We estimated the dependence of $T_{\rm c}$ from two different criteria, $T_{\rm c}^{\rm onset}$ and $T_{\rm c}^{\rm offset}$ corresponding to a drop from the normal state value of 1\,\% and 99\,\% of the full transition respectively. Both appear in the $p\!-\!T$ phase diagram Fig.\,\ref{fig:Tc_vs_P}, and indicate that, above 2\,GPa, our measurements strongly disagree with the susceptibility study reported by Thomas \textit{et al.}\citep{Thomas1996}. Although, up to 4.38\,GPa, a well-defined sharp transition with a pressure independent amplitude is observed, at higher pressure it is not trivial to define $T_{\rm c}^{\rm onset}$ and the full transition could happen to some degree at slightly lower temperature; at 5.12\,GPa the transition is not complete and only $T_{\rm c}^{\rm onset}$ is defined.\\ \indent
An example of data sets from the four physical properties at selected pressures (0.34, 2.62 and 4.38\,GPa) is given in Fig.\,\ref{fig:TEP_CAC_Susc_Res} to illustrate the concurrence of the results. At each pressure, clean and sharp transitions observed in the heat capacity and ac-magnetic susceptibility take place simultaneously. These anomalies are directly related to the bulk properties of the material and their midpoints temperature which are indicated by vertical dotted lines are located at temperatures very close to $T_{\rm c}^{\rm R=0}$. The heat capacity expressed as $C$/$T$ gives an estimate of the electronic specific heat coefficient $\gamma$ which decreases with pressure increase. The same trend is clearly observed at low pressure (up to 2\,GPa) in the accurate measurements reported in Ref.\,\onlinecite{Lengyel2009}. The amplitude of the anomaly then reaches a maximum near $p_V$ and qualitatively agrees with previous measurements\cite{Holmes2004,Holmes2007}. The discrepancy between our results and other studies is due to the variations in sample quality and experimental setup.\\ \indent
The resistive transitions on the other hand are stretched over $\sim\,0.4-1$\,K at 2.63 and 4.38\,GPa with their $T_{\rm c}^{\rm onset}$ appearing at higher temperatures where strictly no anomaly is seen in either $C_{\rm ac}$ or $\chi_{\rm ac}$ data (inset Fig.\,\ref{fig:TEP_CAC_Susc_Res}). This broadening is intrinsic to \CeCuSi{} for which a phenomenon of filamentary superconductivity drives the resistance property\cite{Holmes2004,Holmes2007} in the intermediate pressure regime. Deviations are observed neither in the $C_{\rm ac}$ nor in $\chi_{\rm ac}$ data until the resistivity values drop by at least 80\,\% which suggest that only a negligible sample volume engages in the filamentary behaviour.\\ \indent
The measured thermoelectric power is typical of Ce compounds close to a magnetic instability; at 2.5\,K we observe values of -12, -2 and 9\,$\mu$V.K$^{-1}$ at 0.34, 2.62 and 4.38\,GPa respectively with a sign change at $p\sim$\,3.5\,GPa as formerly documented in Ref.\,\onlinecite{Seyfarth2012,Jaccard1985,Zlatic2003}. $S$ is also dominated by the development of filamentary superconductivity hence the concurrence of $\rho$ and $S$ superconducting transition onset at each pressure. The $S\sim\,0$ values obtained in the superconducting state are expected and confirm the good functioning of the technique.\\ \indent
In Fig.\,\ref{fig:Tc_vs_P}, $T_{\rm c}$ values obtained from $\chi_{\rm ac}$, $\rho$ and $C_{\rm ac}$ (both onset and offset criteria for $\chi_{\rm ac}$ and $\rho$) are plotted together with those from Thomas \textit{et al.}\cite{Thomas1996} $\chi_{\rm ac}$ data. Between 0.35 and 1.72\,GPa the resistivity $T_{\rm c}^{\rm onset}$ interpolation is based on previous observations\cite{Vargoz1998,Holmes}. Susceptibility and heat capacity data clearly indicate the sample bulk properties and despite a slight pressure related broadening of the superconducting transitions in $\chi_{\rm ac}$, at each pressure, they correspond to $T_{\rm c}^{\rm R=0}$ and pinpoint the collapse of superconductivity at around 5-5.5\,GPa. In contrast, the previously published susceptibility data display a maximum $T_{\rm c}$ at 3.1\,GPa materialised by a significant change of slope and suggest the persistence of superconductivity up to 9.5\,GPa ($T_{\rm c}$ = 0.9\,K, not shown). While in some samples\cite{Holmes2005} tiny resistivity drops have been observed in a similar pressure range no evidence of superconductivity was ever detected in bulk measurements above $\sim$\,5.5\,GPa down to 0.5\,K.\\ \indent
\section{Discussion}
The major attributes of the present sample $\rho(T,p)$ conform to previous reports\cite{Holmes2004,Seyfarth2012,Bellarbi1984}. Among them, $\rho(T)$ at high temperatures and $T_{max}^\rho(p)$ increase steadily with $p$, while $\rho_0(p)$, $A(p)$ and $n(p)$ obtained from a fit of $\rho=\rho_0+AT^n$ to low temperature data above $T_{\rm c}$ support the results from Ref.\,\onlinecite{Seyfarth2012}. Despite being a cut out of the single crystal used in Ref.\,\onlinecite{Seyfarth2012} and despite using an identical setup in terms of electrical connection and pressure medium, the $\rho_0$(0.34\,GPa)\,=\,1.15\,$\mu\Omega$.cm and $\rho_0$(4.38\,GPa)\,=\,10.7\,$\mu\Omega$.cm appear to be three times greater than that reported in the previous study. Moreover, we observed from the signatures in $C_{\rm ac}$, $\chi_{\rm ac}$ and $\rho$ a maximum superconducting $T_{\rm c}\sim$\,2\,K lower than that reported in Ref.\,\onlinecite{Seyfarth2012} where $T_{\rm c}\sim$\,2.5\,K. These results are coherent with a \CeCuSi{} sample of somewhat lower quality. In fact our results seem more concordant with the results obtained by Holmes \textit{et al.}\cite{Holmes2004} which were collected on a crystal from a different batch and in a helium loaded diamond anvil cell. It is for such delicate and generally destructive studies where reproducibility, due to $p$ conditions and to the sample itself, is an issue that the single setup multiprobe approach on a unique sample is extremely valuable.\\ \indent
The weak pressure dependence of $T_{\rm c}$ observed in the bulk properties between 0.3 and 0.8\,GPa is consistent with the data presented in Ref.\,\onlinecite{Lengyel2011}. Further comparison with Ref.\,\onlinecite{Yuan2003} is difficult because these measurements have been performed on partially Ge-substituted samples which results in modest maximum of the resistive $T_{\rm c}$ ($\sim$\,0.95\,K) caused by the pair breaking effect of nonmagnetic disorder in accordance with a large residual resistivity (see also Ref.\,\onlinecite{Ren2015}).\\ \indent
In the high sensitivity magnetic susceptibility measurements we note the absence of any feature at temperatures around the resistivity $T_{\rm c}^{\rm onset}$. It implies that the broadening of the superconducting transition in the intermediate pressure regime (1.5-3\,GPa) results from a minute part of the sample involved in some form of filamentary superconductivity. This observation has been previously investigated by Holmes \textit{et al.}\cite{Holmes2007} who showed that when exciding the critical current density, the high temperature part of the transition disappears and a sharp transition is recovered. The configuration of our setup was not suited to such tests but the susceptibility results lead to a similar conclusion.\\ \indent
\\ \indent 
\section{Conclusion}
We have developed a multiprobe setup which enabled the investigation of four different physical quantities, simultaneously for some of them, on a unique sample at extremes of pressure and temperature. $S$, $C_{\rm ac}$, $\chi_{\rm ac}$ and $\rho$ measurements carried out on the high quality \CeCuSi{} single crystal provide directly comparable datasets. The $p\!-\!T$ phase diagram derived from these measurements (Fig.\,\ref{fig:Tc_vs_P}) reveals a clear discrepancy with the previous high pressure magnetic study presented in Ref.\,\onlinecite{Thomas1996}. Furthermore, our results verify the $p\!-\!T$ phase diagram proposed in Ref.\,\onlinecite{Seyfarth2012} and exposed the surface or filamentary nature of the resistive superconducting transition broadening at intermediate pressures. The temperature of the sharp transitions obtained from $\chi_{\rm ac}$ measurements are in good agreement with the $C_{\rm ac}$ results which confirm the bulk origin of superconductivity, and identify the most reliable criterion for defining $T_{\rm c}$ at $T_{\rm c}^{\rm R=0}$.
\begin{acknowledgments}
We acknowledge G. Seyfarth for his numerous comments and suggestions, the technical assistance from M. Lopes, and the financial support from the Swiss National Science Foundation through Grant No. 200020-137519. Pablo Pedrazzini is a member of CONICET. 
\end{acknowledgments}
\bibliography{SSC_CeCu2Si2}
\end{document}